\documentclass[a4paper]{jpconf}

\usepackage
{hyperref}
\usepackage{amsthm}
\usepackage{amsmath}
\usepackage{amssymb}
\usepackage{a4wide}

\usepackage{tikz} 
\usepackage{caption} 
\usepackage{iopams}

\newcommand{\mb}[1]{\ensuremath{\mathbb{#1}}}

\newcommand{\R}{\mb{R}}

\newcommand{\sgn}{\mathop{\mathrm{sgn}}}

\renewcommand{\d}{\ensuremath{\partial}}

\newfont{\bl}{msbm10 scaled \magstep2}

\newcommand{\beq}{\begin{equation}}
\newcommand{\eeq}{\end{equation}}

\newcommand{\notmid}{\mid\kern-0.5em\not\kern0.5em}

\newcommand{\eps}{\varepsilon}

\newenvironment{pr}{\begin{proof}[\textbf{Proof:}] \ }{\end{proof}}
\theoremstyle{definition}
\newtheorem{thm}{Theorem}[section]
\newtheorem{lem}[thm]{Lemma}
\newtheorem{prop}[thm]{Proposition}
\newtheorem{ex}[thm]{Example}
\newtheorem{cor}[thm]{Corollary}

\renewcommand{\d}{{\mathrm{d}}}

\newcommand{\ep}{\varepsilon}
\newcommand{\cN}{{\mathcal N}}
\newcommand{\cD}{{\mathcal D}}

\begin{document}
  
\title{On geodesics in low regularity}
\author{Clemens S\"amann and Roland Steinbauer}
\address{\textsc{Faculty of Mathematics, University of Vienna, Austria}}
\ead{clemens.saemann@univie.ac.at, roland.steinbauer@univie.ac.at}

\begin{abstract}
We consider geodesics in both Riemannian and Lorentzian manifolds with metrics 
of low regularity. We discuss existence of extremal curves for continuous 
metrics and present several old and new examples that highlight their subtle 
interrelation with solutions of the geodesic equations. Then we turn to the 
initial value problem for geodesics for locally Lipschitz continuous 
metrics and generalize recent results on existence, regularity and uniqueness 
of solutions in the sense of Filippov.
\vskip 1em
  
\noindent
\emph{Keywords:} geodesics, extremal curves, low regularity, Filippov solutions 
\medskip
    
\noindent 
\emph{MSC2010:} 
53B30, 
53C22,  
34A36, 
83C99  
    
\end{abstract}

\section{Introduction}

Recently there has been an increased interest in several low regularity aspects 
of general relativity and of semi-Riemannian geometry. In particular, it has 
been confirmed that the regularity class of $C^{1,1}$ (the metric possessing 
locally Lipschitz continuous first derivatives) can rightly be seen as the 
threshold of classical theory. In addition to the well-known fact that the 
geodesic equation can be locally uniquely solved by classical ODE-theory, the 
exponential map retains maximal regularity and 
convex neighborhoods exist \cite{Min:15,KSS:14}. In the Lorentzian case 
the bulk of causality theory remains valid \cite{CG:12,Min:15,KSSV:14} 
as well as the classical singularity theorems \cite{KSSV:15,KSV:15,GGKS:17}. On 
the other hand studies of geometries of regularity below $C^{1,1}$ (see e.g.\ 
\cite{CG:12,Sae:16,S:15,GLS:17}) have revealed important 
differences in many of the well known concepts and have emphasized 
the significance of regularity issues.

In this contribution we focus on geodesics in low regularity in Riemannian as 
well as in Lorentzian manifolds. This topic again is of particular interest 
since many of the facts well known for smooth metrics fail to extend to a 
regularity below $C^{1,1}$. For example in the classical setting the unique 
local solutions of the geodesic equation are locally extremal curves for the 
length functional. Conversely extremal curves are pregeodesics. Moreover 
Lorentzian geodesics and maximal curves have a causal character.

In case of insufficient regularity  
the question of existence of extremal curves becomes a separate issue. It has 
long been answered affirmatively in the case of continuous Riemannian metrics 
and we will discuss the recent equally positive answer in the Lorentzian case. 
It might be surprising that already in cases where the geodesic equation is 
(classically locally) solvable but not uniquely so (in particular, for metrics 
of H\"older class $C^{1,\alpha}$ for any $\alpha<1$) the connection between its 
solutions and extremal curves becomes subtle. We will discuss some classical and 
some new examples in that realm at some length. We will supplement them by 
examples (again old and new) which demonstrate the failure of some of the usual 
causality properties in Lorentzian manifolds.  

Complementing this line of investigation we will transfer some recent results on 
solutions of the geodesic equation for impulsive gravitational wave spacetimes 
into a more abstract setting. In particular, we will discuss existence and 
regularity of solutions to the geodesic equation for locally Lipschitz 
semi-Riemannian metrics using the solution concept of Filippov \cite{Fil:88} 
for ODEs with discontinuous right hand side. We also present some sufficient 
conditions for uniqueness, which have proved to be useful in the context of 
applications. Finally we provide an outlook to open questions and 
related problems. 
\medskip

We end this introduction by fixing some notions and notations. All manifolds 
will be assumed to be of class $C^\infty$  (which is no loss of generality, 
\cite[Thm.\ 2.9]{Hirsch}) and we will only lower the regularity of the metric. 
Hence e.g.\ by a continuous spacetime $(M,g)$ we will mean a smooth, connected 
manifold $M$ of dimension $n\geq 2$ equipped with a continuous Lorentzian metric 
$g$ with a time orientation induced by a (continuous) timelike vector field. Our 
notations are quite standard and we generally follow \cite{ON}. Mindful of the 
above discussion a geodesic will always mean a solution (of some sort) of the 
geodesic equation and should be strictly distinguished from extremal curves.

\section{Extremal curves}
As indicated in the introduction, below a regularity of $C^{1,1}$ the 
notions of extremal curves and geodesics no longer coincide. We discuss 
existence of extremal curves for continuous metrics and study their relation to 
geodesics. We start with the case of Riemannian metrics.

\subsection{The Riemannian case}\label{subsec:rc}

It is a classical result by Hilbert \cite{Hil:04} (using the Theorem of 
Arzela-Ascoli) that for \emph{con\-ti\-nuous} Riemannian metrics minimizing curves 
always exist locally. The global existence of minimizing curves is a corollary 
of the Hopf-Rinow-Cohn-Vossen Theorem for length spaces \cite[Thm.\ 
2.5.28]{BBI:01}. To be precise, by \cite{Bur:15}, a continuous Riemannian metric 
gives rise to a length space, and hence if the corresponding metric space is 
complete any two points can be connected via a minimizing curve.

So the situation of existence of minimizing curves for continuous Riemannian 
metrics is exactly as in the smooth case. However, the relation between such 
minimizing curves and geodesics turns out to differ drastically from the smooth 
situation if the regularity of the metric drops below $C^{1,1}$. 

Indeed classical examples show the explicit failure of the initial
value problem for geodesics to be \emph{uniquely} solvable for $g\in 
C^{1,\alpha}$ for any fixed $\alpha<1$ \cite{Har:50}, 
while it is possible that at the same time all the usual properties hold 
\cite{HW:51}: minimizing curves are locally unique, the boundary value problem 
for geodesics is locally uniquely solvable and even in `singular points' there 
is a locally minimizing curve starting off in any direction.

However, another classical example \cite{HW:51} shows that for $g\in 
C^{1,\alpha}$ again for any fixed $\alpha<1$ geodesics need not be 
even minimizing locally. This example also shows that again even locally the 
boundary value problem for geodesics is non-uniquely solvable. As the same 
phenomenon occurs also in the Lorentzian case by a modification of the original 
example we will discuss it in some detail below.

On the positive side the regularity of minimizing curves is slightly better than 
one would expect. A classical result in this  direction is the 
following: If the metric is $C^1$, then minimizing curves are actually geodesics 
and are of regularity $C^2$. This can be seen by using the trick of Du 
Bois-Reymond \cite[Ch.\ I, \textsection 6]{Bol:04}, which we briefly recall 
here.

Let $g$ be a $C^1$-Riemannian metric. Since it suffices to argue locally 
we consider $\gamma\colon[a,b]\rightarrow \R^n$,  
a locally Lipschitz continuous and minimizing curve from $p=\gamma(a)$ to 
$q=\gamma(b)$. We parametrize $\gamma$ by $g$-arclength, i.e., 
$g_{\gamma(t)}(\dot\gamma(t),\dot\gamma(t))=1$ almost everywhere. That $\gamma$ is 
minimizing implies that
\begin{equation*}
 \left. \frac{d}{d \eps}\right\rvert_{\eps=0} L_g(\gamma+\eps \phi) = 0\,,
\end{equation*}
where $L_g$ is the length functional of $g$, i.e., $L_g(\gamma) = \int_a^b 
F(t,\gamma(t),\dot\gamma(t))\,\d t$ with 
$F(t,\gamma(t),\dot\gamma(t))$$=\sqrt{g_{\gamma(t)}(\dot\gamma(t),\dot\gamma(t))}$, and 
moreover $\phi$ is a smooth function with compact support in $[a,b]$. For 
simplicity we calculate the following in one dimension, however the general case 
poses no additional difficulties. Integration by parts yields
\begin{equation}\label{eq:ibp}
 0 = \left.\frac{d}{d \eps}\right\rvert_{\eps=0} L_g(\gamma+\eps \phi) = \int_a^b F_{\gamma} \phi + F_{\dot\gamma} \phi'\,\d t 
= \int_a^b (F_{\dot\gamma} - \int_a^t F_{\gamma}\,\d s) \phi'\,\d t\,.
\end{equation}
Thus, since $\phi$ is arbitrary, $F_{\dot\gamma}-\int_a^t F_{\gamma}\,\d s$ 
is constant and in the general higher dimensional case this yields that
\begin{equation}\label{eq:dBR-c}
 F_{\dot\gamma^i} - \int_a^t F_{\gamma^i}\,\d s = c_i = \text{constant}\,,
\end{equation}
for $i=1,\ldots,n$. Note that $\gamma$ minimizes on any subinterval $[a,s]$ 
($s\leq b$), and hence we obtain from \eqref{eq:dBR-c} in coordinates $(x^i)$ 
($\gamma^i = x^i\circ \gamma$) that 
\begin{equation*}
 g_{ij}\, \dot\gamma^j - \frac{1}{2}\int_a^s \frac{\partial g_{lj}}{\partial 
x^i}\dot\gamma^l\dot\gamma^j\,\d r  = c_i\,,
\end{equation*}
and by multiplying with $g^{-1}=(g^{ij})$ this gives
\begin{equation}\label{eq:dBR-reg}
 \dot\gamma^m = g^{im} \left(\frac{1}{2}\int_a^s \frac{\partial g_{lj}}{\partial 
x^i}\dot\gamma^l\dot\gamma^j\,\d r +  c_i\right)\,.
\end{equation}
The right-hand-side of \eqref{eq:dBR-reg} is continuous because 
$\frac{\partial g_{lj}}{\partial x^i}$ is continuous and $\dot\gamma^i$ is bounded. 
Thus $\gamma$ is $C^1$ and using this information one sees again by 
\eqref{eq:dBR-reg} that $\gamma$ is in fact $C^2$.

Finally it follows that $\gamma$ actually is a geodesic by the usual argument 
using integration by parts on the other term in \eqref{eq:ibp}. 
\medskip

A recent and enhanced regularity result is given in \cite{LY:06}: If the 
Riemannian metric is of H\"older-regularity $C^{0,\alpha}$ with $0<\alpha\leq 
1$, then minimizing curves are of regularity $C^{1,\beta}$, where $\beta = 
\frac{\alpha}{2-\alpha}$. This regularity $C^{1,\beta}$ is optimal as there are 
examples of minimizers with respect to a $C^{0,\alpha}$ Riemannian metric on 
$\R^2$ that are not $C^{1,l}$ for $l>\beta$ \cite[Thm.\ 1.1]{LY:06}
\medskip

Finally, we review a key example of Hartman and Wintner given in \cite{HW:51}.
\begin{ex}(The Hartman-Wintner example \cite[Sec.\ 5]{HW:51})\label{ex:hw}
We consider $M:=(-1,1)\times\R$ with the metric 
\begin{equation}\label{eq:hwm}
 g_{(x,y)} =  dx^2 + (1-|x|^\lambda)dy^2\,,
\end{equation}
where $1<\lambda < 2$. The metric is of 
H\"older-regularity $C^{1,\lambda-1}$ 
and smooth off the $y$-axis. The geodesic equations are
\begin{align}
 x'' + \frac{\lambda}{2} |x|^{\lambda-1}\sgn{x} (y')^2 = 0,\label{eq-HW-geo}
 \qquad
 y'' -\frac{\lambda}{2} \frac{|x|^{\lambda-1}\sgn(x)}{1-|x|^{\lambda}}x'y' = 
0\,.
\end{align}
Moreover, it suffices to consider only the initial value 
$(x_0,y_0)=(0,0)$ since the metric does not 
depend on $y$ at all. Note that since the metric is $C^1$,  
$g_{\gamma(s)}(\dot\gamma(s),\dot\gamma(s))$ is constant and we 
parametrize any geodesic by arclength.
The $y$-equation in \eqref{eq-HW-geo} is equivalent to  $(1-|x|^\lambda)y' = c =$ constant.
In case $c=0$ we have $y=0$, so the $x$-equation in \eqref{eq-HW-geo} is 
trivial and the geodesic is simply given by $y=0$.
If $c\neq 0$, then $y'\neq 0$ and $y$ is strictly monotonous along any such 
geodesic. Parametrizing by arclength gives 
\begin{equation}\label{eq-HW-xcr}
 x' = \pm \sqrt{1 - \frac{c^2}{1 - |x|^{\lambda}}}\,,
\end{equation}
with $c\in[-1,1]$. If $c^2=1$, then $x=0$ is the only 
solution to \eqref{eq-HW-xcr} and we denote this geodesic by $\gamma_0$. 
If $c^2<1$, then the right-hand-side of 
\eqref{eq-HW-xcr} is (initially) $C^1$ and thus there is a unique solution 
with initial condition $x(0)=0$. Given this unique solution $x$, we can 
uniquely solve the $y$-equation by integrating, i.e., $y(s) = \int_{0}^s \frac{c}{1 - |x(r)|^\lambda} \d r$.
To summarize, this means that at $(0,0)$ for every initial direction there is a 
unique geodesic, and by symmetry the initial-value-problem is uniquely 
solvable for arbitrary data. 

To determine the shape of the geodesics set 
$c=\pm\sqrt{1-\eps}$ for $\eps\in[0,1]$ and denote by $\gamma_{\pm\eps}$ the 
geodesics starting at $(0,0)$ with initial velocity 
$(\pm\sqrt{\eps},\sqrt{1-\eps})$, 
see Figure~\ref{fig:hw}.
A short calculation shows that $\gamma_\eps$ reaches $x=\sqrt[\lambda]{\eps}$ 
in finite time. We denote the corresponding parameter value by $s_0$ and the 
$y$-value by $y_1$, i.e., $\gamma_\eps(s_0)=(\sqrt[\lambda]{\eps},y_1)$. Then 
$\dot\gamma_\eps(s_0)=(0,1/\sqrt{1-\eps})$ is vertical and since $g$ is 
independent of $y$ we can reflect $\gamma_\eps$ at $y=y_1$. Thus by symmetry 
$(0,0)$ is connected to $(0,2 y_1)$ by three distinct geodesics: 
$\gamma_{\pm\eps}$ and $\gamma_0$. At this point we vary $\eps$ and it is not hard to see that 
$y_1(\eps)\to 0$ as $\eps\searrow 0$. This implies that the boundary value 
problem for geodesics is not uniquely solvable in any neighborhood of any 
point on the $y$-axis. 

Finally, it is again not hard to see that the geodesic $\gamma_0$ is not 
minimizing between any of its points and that in fact $\gamma_{\pm\eps}$ are 
minimizing between its endpoints. From here we obtain that even locally 
minimizing curves are not unique and there is no minimizing curve with initial 
velocity $(0,1)$.

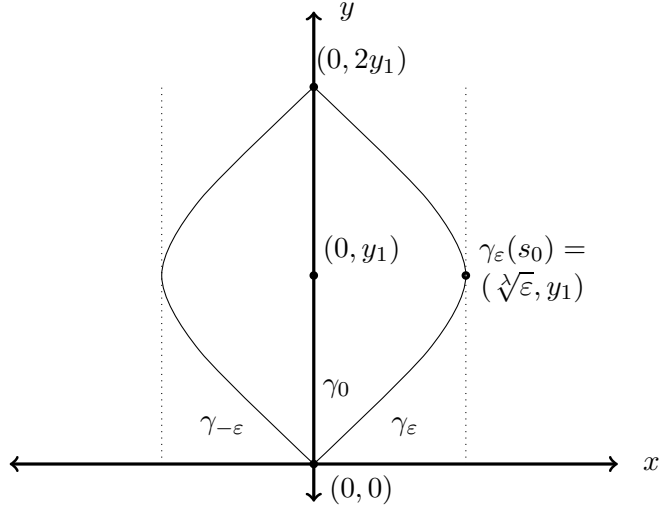
\begin{figure}[h]\centering
    \begin{tikzpicture}
   \draw [<->,very thick] (-4,0) -- (4,0) node[align=right] {$\qquad x$};
   \draw [<->,very thick] (0,-.5) -- (0,6) node[align=right] {$\qquad y$};
   \draw [very thick] (0,0) circle (1pt) node[align=right,below] 
    {$\qquad\quad (0,0)$};
    \draw plot [smooth] coordinates {(0,0) (1.5,1.5) (2,2.5) (1.5,3.5) (0,5)};
    \draw [dotted] (2,0) -- (2,5);
    \draw (1.2,.5) node {$\gamma_\eps$};
   
\draw plot [smooth] coordinates {(0,0) (-1.5,1.5) (-2,2.5) (-1.5,3.5) (0,5)};
    \draw [dotted] (-2,0) -- (-2,5);
    \draw (-1.2,.5) node {$\gamma_{-\eps}$};

\draw [very thick] (2,2.5) circle (1pt) node[align=right,above] 
    {$\qquad\qquad \gamma_\eps(s_0)=$};
\draw [very thick] (2,2) node[align=right,above] 
    {$\qquad\qquad (\sqrt[\lambda]{\eps},y_1)$}; 
\draw [very thick] (0,5) circle (1pt) node[align=right,above] 
    {$\qquad\quad (0,2y_1)$};
\draw [very thick] (0,2.5) circle (1pt) node[align=right,above] 
    {$\qquad\quad (0,y_1)$};
\draw [very thick] (.3,1) node {$\gamma_0$};

  \end{tikzpicture}
  \caption{The minimizing geodesics $\gamma_{\pm\eps}$ and the 
   non-minimizing one $\gamma_0$ for the metric \eqref{eq:hwm}.} 
\label{fig:hw}
\end{figure}
\end{ex}

\subsection{The Lorentzian case}\label{sec-lor}
In complete analogy with the Riemannian case local existence of maximizing 
causal curves holds for \emph{continuous} Lorentzian metrics. The key notion 
here is \emph{global hyperbolicity}, which for continuous metrics has been 
studied in \cite{Sae:16}. There a spacetime is called globally hyperbolic if it 
is non-totally imprisoning and the causal diamonds are compact. Recall that a 
\emph{Cauchy hypersurface} is a set $S\subseteq M$ that is met exactly once by 
any inextendible (locally Lipschitz) causal curve. The existence of a Cauchy hypersurface is equivalent to 
global hyperbolicity also in this low regularity \cite[Thm.\ 5.7 and Thm.\ 
5.9]{Sae:16}. We may establish without much effort:

\begin{thm}(Existence of locally maximizing curves)\label{thm:lm}
 Every point $p$ in a continuous spacetime $(M,g)$ possesses a neighborhood
 $U$ such that any two $U$-causally related points can be joined by a 
 maximizing (in $U$) causal curve.
\end{thm}

\begin{pr}
Let $\hat{g}$ be a smooth Lorentzian metric on $M$, with the property that 
the lightcones of $g$ are contained in the timelike cones of $\hat{g}$ (see  
\cite{CG:12}). 
Let $p\in M$ then there exists a 
$\hat{g}$-globally hyperbolic neighborhood $(U, x^\mu)$ of $p$ by \cite[Thm.\ 
2.14]{MS:08}, i.e., in the $x^\mu$-coordinates one has that $x^0=0$ is a Cauchy 
hypersurface in $U$ with respect to $\hat{g}$. Then obviously this is also a 
Cauchy hypersurface with respect to $g$. Hence $(U,g\rvert_U)$ is globally 
hyperbolic by \cite[Thm.\ 5.7]{Sae:16} and thus maximal (in $U$) causal curves 
exist between any two (in $U$) causally related points by  \cite[Prop.\ 
6.4]{Sae:16}.
\end{pr}

Here we have already made use of the Avez-Seifert result for continuous metrics 
which we quote below as the global analogue of the above result. Note, however, 
that Theorem \ref{thm:lm} can be established without using the global result: 
one may just use that there are local bounds on the arclength of causal curves 
with respect to a complete Riemannian metric and then use the Theorem of 
Arzela-Ascoli in the form of the limit curve theorem \cite{Min:08,CG:12,Sae:16}.

\begin{thm}(Avez-Seifert, \cite[Prop.\ 6.4]{Sae:16})
  Let $(M, g)$ be a globally hyperbolic continuous spacetime. Then there is a 
  maximal causal curve connecting any two points in the spacetime which 
  are causally related. 
\end{thm}

We summarize the Riemannian and Lorentzian results on the existence of minimal 
and maximal curves in the following table. 
\begin{center}
  \begin{tabular}{l|l|l}
    $g\in C^0$&local&global\\[.2em]\hline &\\[-.7em]
    
    Riemannian & length sructure, & Hopf-Rinow-Cohn-Vossen \\
    &Arzela-Ascoli \cite{Hil:04} & in complete length spaces 
    \cite[Thm.\ 2.5.28]{BBI:01}, \cite{Bur:15} 
    \\ \hline &\\[-.7em]
    Lorentzian & existence of globally &
    continuous Avez-Seifert \cite{Sae:16} \\ 
    &hyperoblic neighborhoods
    &
  \end{tabular}
\end{center}
\medskip

Again as in the Riemannian case the relation of maximizing causal curves with 
geodesics becomes subtle for a regularity below $g\in C^{1,1}$. Indeed Example 
\ref{ex:hw} can be modified to yield a Lorentzian counterexample.

\begin{ex}(The Lorentzian Hartman-Wintner example)\label{ro:hwl}
We consider $M:=\R\times(-1,1)\times\R$ with the $C^{1,\lambda-1}$-metric 
\begin{equation*}
 g_{(t,x,y)} =  -dt^2+ dx^2 + (1-|x|^\lambda)dy^2\,,
\end{equation*}
where again $1<\lambda < 2$. Then the geodesic equations for $x$ and $y$ are 
just equations \eqref{eq-HW-geo} and in addition we have the trivial equation 
$t''=0$. 

Again it suffices to consider geodesics starting at the origin. By Example 
\ref{ex:hw} the initial value problem is uniquely solvable and we again consider 
some special solutions. Defining $y_1$ as in Example \ref{ex:hw} we consider the 
timelike geodesic $\Gamma_0(r)=(2\sqrt{2} s_0 r,0,2 y_1 r)$ ($r\in[0,1])$. To 
see that $\Gamma_0$ is timelike, note that $s_0 < y_1 < 2 s_0$. Moreover 
$L(\Gamma_0) = \sqrt{8s_0^2 - 4 y_1^2} < 2 s_0$. On the other hand set 
$\Gamma_{\pm\eps}(s)=(\sqrt{2}s,\gamma_{\pm\eps}(s))$ ($s\in[0,2 s_0]$), where 
$\gamma_{\pm\eps}$ are the minimizing geodesics of Example \ref{ex:hw}. Then the 
geodesics $\Gamma_{\pm\eps}$ are normalized to eigentime and so the lengths 
$L(\Gamma_{\pm\eps}) = 2 s_0$ coincide with parameter length. Therefore we 
obtain without effort similar results as in the original example.

Indeed $\Gamma_{\pm\eps}(s_0)=(\sqrt{2}s_0,\pm\sqrt[\lambda]{\eps},y_1)$.
Then the points $(0,0,0)$ and $(2\sqrt{2}s_0,0,2y_1)$ are connected by the
three distinct geodesics $\Gamma_{\pm\eps}$ and $\Gamma_0$ and again the 
boundary value problem for geodesics is not uniquely solvable. 
Moreover, $\Gamma_0$ is not maximizing between any of its points since the timelike 
curves $\Gamma_{\pm\eps}$ are longer: $L(\Gamma_0) < 2 s_0 = L(\Gamma_{\pm 
\eps})$. Finally there is no maximizing curve with initial 
velocity$(\sqrt{2}s_0,0,y_1)$ and by symmetry maximizing curves are not unique.
\end{ex}

This example is complemented by the following one from \cite{CG:12} which, in 
particular, shows that the push-up principle of causality theory fails to hold 
in H\"older-regularity $C^{0,\lambda}$ with $\lambda<1$. Recall that the push-up 
principle states that if two points can be connected via a causal curve that 
contains a timelike segment, then the points can be connected by a timelike 
curve. This H\"older-regularity is optimal as the push-up principle holds for 
(locally) Lipschitz continuous metrics \cite[Cor.\ 1.17 and Prop.\ 1.23]{CG:12}.

\begin{ex}(Bubbling \cite[Ex.\ 1.11]{CG:12})\label{ex:cg}
We consider the manifold $\R^2$ with the metric
\begin{equation}
 g_{(u,x)} = -du^2 + 2 (|u|^\lambda-1)du\, dx + |u|^\lambda(2-|u|^\lambda)dx^2\,,
\end{equation}
where $0<\lambda<1$ and $\partial_u$ gives the time-orientation. The metric is 
$\lambda$-H\"older regular and smooth of the $x$-axis. Null curves branch off 
from the $x$-axis and points in the region between the first branching null 
curve and the $x$-axis (which is also null) can be connected to the origin via 
causal curves (even of positive length) but not by timelike curves, see 
\cite[Fig.\ 1.1]{CG:12}. This region is called the \emph{causal bubble} and this demonstrates the failure of the 
push-up 
principle.

\end{ex}

A slight modification of the above example yields the existence of a maximal 
causal curve in a $C^{0,\lambda}$-spacetime which does not have a causal 
character. This example can be found in detail in \cite[Cor.\ 5.5]{KS:17}.
    
\begin{ex}\label{ex:ks}
In the above Example \ref{ex:cg} we put for simplicity $\lambda=\frac{1}{2}$ and 
consider only the manifold $M:=(-1,1)\times\R$. Let $q:=(u_0,x_0)$ be in the 
(upper right) \emph{bubble region}, i.e., $0 
< u_0 < \min(\frac{x_0}{4},1)$. As this spacetime is strongly causal and the 
causal diamond of the origin and $q$ is compact, there is a maximal causal curve 
from the origin to $q$. This maximal causal curve cannot have a causal 
character: The curve has to first go along the $x$-axis, where it is null and 
after it leaves the $x$-axis it is in the smooth part of the spacetime. 
Moreover, there it has to have positive length and thus must be timelike.
\end{ex}


Finally, we remark that the Du Bois-Reymond trick does not work in the 
Lorentzian setting. If $g$ is a $C^1$-Lorentzian metric and $\gamma$ is a 
maximal causal curve, then it is not clear that $\gamma$ has to be timelike 
even if $\gamma$ has positive length. Therefore, one cannot ensure that the 
variation $\gamma+\eps \phi$ (compare the discussion the Riemannian case in 
Section \ref{subsec:rc}) is causal, hence there is no reason that 
$\left.\frac{d}{d \eps}\right\rvert_{\eps=0} L_g(\gamma+\eps \phi) = 0$.
The argument only works out in case we already know that the maximizing curve 
is timelike. 

\section{The geodesic equation for locally Lipschitz metrics}
In this section we deal with the geodesic equation for semi-Riemannian 
manifolds $(M,g)$ with $g\in C^{0,1}$.
%
%
The motivation for studying this regularity class comes on the one hand from 
the fact that here some crucial aspects of causality theory remain valid, cf.\ 
\cite{CG:12} and Section~\ref{sec-lor} above,  
and on the other hand that relevant exact solutions of general relativity such 
as impulsive gravitational wave spacetimes have this regularity.

Recall that in  this regularity class the right-hand-side of the geodesic 
equation is only guaranteed to be locally bounded but \emph{not} to be 
continuous. It turns out that a fruitful way to deal with these `ODEs with 
discontinuous right hand side' is to use the Filippov solution concept 
\cite{Fil:88}, which we will now briefly review (see \cite{Cor:08} for 
an application-driven introduction).

\subsection{Filippov solutions}
The key idea of Filippov's approach is to replace an ODE 
\begin{equation}\label{eq:ode}
  \dot x(t) = f(x(t))\qquad (t\in I)\,,
\end{equation}
($I$ is some interval, $f\colon \R^n\supseteq D\rightarrow \R^n$
possibly of low regularity) by a \emph{differential inclusion}
\begin{equation}\label{eq:di}
  \dot x(t)\in{\mathcal F}[f](x(t)),
\end{equation}
where the \emph{Filippov set-valued map} ${\mathcal F}[f]\colon 
D\to {\mathcal K}_0(\R^n)$ (the collection of all nonempty,
closed and convex subsets of $\R^n$) associated with $F$ is defined as
its essential convex hull
\begin{equation}\label{eq:fsvm}
{\mathcal
  F}[f](x):=\bigcap_{\delta>0}\bigcap_{\mu(S)=0}
{\text{co\,}}\Big(f(B(x,\delta)\setminus S)\Big).
\end{equation}
Here $\text{co}$ denotes the closed convex hull, $B(x,\delta)$
is the closed ball of radius $\delta$ centered at $x$, and $\mu$ is the Lebesgue
measure, cf.\ \cite[Sec.\ 2]{PR:97}.

The key idea is to look at the set of values $f$ takes near a 
point of discontinuity. This fact lies also at the heart of the compatibility 
of this approach with approaches based on regularization, see \cite[Sec.\ 
3.3]{Hal:08}. 
The explicit calculation of a Filippov set-valued map can be non-trivial, 
however, there exists a calculus to at least bound this set \cite{PS:87}. Also 
note that a Filippov set-valued map is actually multi-valued only at the points 
of discontinuity of $f$. 
\medskip

Now a \emph{Filippov solution} of the ODE \eqref{eq:ode} is defined to be an
absolutely continuous curve $x\colon J\to U$, defined on some interval 
$J\subseteq I$, that satisfies the inclusion relation \eqref{eq:di} almost 
everywhere. 

Recall that a curve ${x:[a,b]\rightarrow \R^d}$ is said to be \emph{absolutely 
  continuous} if for every ${\ep>0}$ there is a ${\delta>0}$ such that for all 
collections of non-overlapping intervals ${([a_i,b_i])_{i=1}^m}$ in ${[a,b]}$ 
with ${\sum_{i=1}^m(b_i-a_i)<\delta}$ we have that ${\sum_{i=1}^m 
  \|x(b_i)-x(a_i)\|< \ep}$. Moreover, recall that an absolutely continuous 
curve 
is differentiable almost everywhere.

Of course, if $f$ is continuous, classical ${\mathcal C}^1$-solution and 
Filippov solutions coincide, but the latter exist under much more general 
assumptions on $f$. In fact, Filippov in \cite{Fil:88} has developed a complete 
theory of ordinary differential equations based on this solution concept which 
has been found to be widely applicable e.g.\ in non-smooth mechanics. Here we 
just state the main existence result.
\begin{thm}(\cite[Thm.\ 7.1]{Fil:88})
  The initial value problem
  \begin{equation}
    \dot x(s)\in A(s,x(s))\ \mbox{a.e.},\quad 
    x(t_0)=x_0 \qquad (t_0,x_0)\in I\times\R^n
  \end{equation}
  has an absolutely continuous solution if the set valued map
  $I\times\R^n\ni (t,x)\mapsto A(t,x)$ satisfies
  \begin{enumerate}
    \item[(i)] $t\mapsto A(t,x)$ is Lebesgue measurable on $I$ for all fixed 
    $x$,
    \item[(ii)] $x\mapsto A(t,x)$ is upper semi-continuous for almost all $t$, 
    and
    \item[(iii)] $\sup_{x\in\R^n}|A(t,x)|\leq\beta(t)\in
    L^1_{\mbox{\scriptsize loc}}(I)$ for almost all $t$.
  \end{enumerate}
\end{thm}
Our main interest lies in the following simple consequence, where we denote by 
$L^\infty_{\mbox{\scriptsize loc}}$ the space of all measurable 
and locally essentially bounded functions.
\begin{cor}(\cite[Thm.\ 7.8]{Fil:88})
  If $f\in L^\infty_{\mbox{\scriptsize loc}}(D,\R^n)$, then for each 
  $(t_0,x_0)\in I\times D$ there is a Filippov solution $x$ of \eqref{eq:ode} 
  with 
  $x(t_0)=x_0$.
\end{cor}

\subsection{Existence of geodesics}
This results lends itself to an application to the geodesic 
equations on semi-Riemannian manifolds $M$ with a $C^{0,1}$-metric. 
First observe that the notion of the essential convex hull is  
invariantly defined on $M$. Indeed ${\mathcal F}(f)(\phi(x))={\mathcal 
F}(f\circ\phi)(x)$ for any diffeomorphism $\phi$ since the respective balls 
in \eqref{eq:fsvm} $B(\phi(x),r)$ and $\phi(B(x,r'))$ can be nested.
Now rewriting the geodesic equations as a first order system one locally 
obtains from the Lipschitz property of the metric an equation of the form 
\eqref{eq:ode} with  $f\in L^\infty_{\mbox{\scriptsize loc}}$. Hence we have:
\begin{cor}(\cite[Thm.\ 2]{Ste:14}\label{thm:main})
  Let $(M,g)$ be a smooth manifold with a $\mathcal{C}^{0,1}$-semi-Riemannian
  metric. Then there exist a Filippov solution of the geodesic equation for 
  arbitrary data $p\in M$, $v\in T_pM$. These solutions 
  possess absolutely continuous velocities (hence, in particular, are 
  ${\mathcal C}^1$-curves).
\end{cor}

Observe that such geodesics do not satisfy the geodesic equation in 
the classical sense. We 
only know that for almost all values of an affine parameter we have that 
$\ddot\gamma^i\in{\mathcal F}(-\Gamma^i_{\,jk}\dot\gamma^j\dot\gamma^k)$. 
Consequently, in general the norm of the tangent $|\dot\gamma|$ will not be 
preserved and hence in the Lorentzian case $\gamma$ will not have a global 
causal character. For the same reason we do not know whether these geodesics are 
locally minimizing resp.\ maximizing curves. Also concerning regularity we do 
not know whether the geodesics are $C^{1,1}$, which would match the limiting 
case $\alpha=1=\beta$ in \cite{LY:06}, see Section \ref{subsec:rc} above. In 
fact, we only obtain that the geodesics are $C^1$ with absolutely continuous 
tangent.

\subsection{Uniqueness of geodesics}
Next we turn to uniqueness. Clearly we cannot hope for any general result and we 
will rather be interested in situations where the $C^{0,1}$-metric is actually 
smooth off some hypersurface. This allows us to cover the case of `matched 
spacetimes' and impulsive gravitational wave geometries, see below. Here we 
start with a general discussion of uniqueness of Filippov solutions for 
\eqref{eq:ode}. In principle essentially one-sided Lipschitz conditions, i.e., 
$(f(x)-f(y))^t (x-y)\leq L \|x-y\|_2^2$ for some $L$ and almost all $x,y$, 
provide one-sided uniqueness results (e.g.\ \cite[Prop.\ 4]{Cor:08}) without 
requiring $f$ to be continuous. However, these results are ill-suited in case 
$f$ is smooth off some hypersurface in $\R^n$ since piecewise smooth functions 
generically fail to be essentially one-sided Lipschitz even in cases where the 
corresponding Filippov solutions are unique, cf.\ \cite[p.\ 53]{Cor:08}.    

Hence we follow a different route and start by making the situation precise.
Suppose that $f\in L^\infty_{\mbox{\scriptsize loc}}(D)$ and that the domain 
$D\subseteq\R^n$ is connected and a disjoint union $D=D^-\dot\cup N\dot\cup 
D^+$, where $D^+$, $D^-$ are open sets and their common boundary, $N=\partial 
D^+=\partial D^-$ is a smooth hypersurface, see the figures
below. Now suppose $f\in C^1(D^\pm)$ up to the boundary $N$ 
and denote by $f^\pm$ the extensions of $f|_{D^\pm}$ to the closure 
$\bar {D}^\pm= D^\pm\cup N$. Finally write $f^\pm_N$ for the projections of 
$f^\pm|_N$ on the unit normal $\vec n$ of $N$, pointing from $D^-$ to $D^+$.
We will abbreviate this situation by saying that the locally bounded function 
$f$ is \emph{smooth off the hypersurface $N$}. 
Then basic ODE-theory on the domains $\bar D^\pm$ yields the following:
\begin{lem}(Simple conditions for uniqueness,
  \cite[Lem.\ 10.2]{Fil:88})\label{lem:uni} Let  $f\in 
L^\infty_{\mbox{\scriptsize 
      loc}}(D,\R^n)$ be smooth off the hypersurface\footnote{In the sense 
    defined above.} $N$. If for $x_0\in N$ we have $f^+_N(x_0)>0$, then in $\bar 
  D^+$ 
  there exists a unique $C^2$-solution of \eqref{eq:ode} starting at $x_0$. 
  Analogous assertions hold for $\bar D^-$ and $f^-_N(x_0)<0$.
\end{lem}
This basic observation also 
leads to the following conclusions: If at a point $x_0\in N$ we have that 
$f^+$ points into $D^+$ ($f^+_N>0$) and $f^-$ points into $D^-$ 
($f^-_N<0$) then there are $C^2$-solutions which proceed into $D^+$ and such 
that proceed into $D^-$ and one speaks of \emph{repulsive trajectories}, see 
Figure \ref{fig:unique3}. Clearly in this case uniqueness of Filippov solutions 
on $D$ fails.

In all other possible cases uniqueness of Filippov solutions can be 
secured (\cite[Prop.\ 5]{Cor:08}). However, if at $x_0\in N$ we have that $f^+$ 
points into $D^-$ ($f^+_N<0$) and $f^-$ points into $D^+$ ($f^-_N>0$) then the 
$C^2$-solutions from either side may be trapped in $N$, a situation 
referred to as \emph{sliding motion}, see Figure \ref{fig:unique4}. 
We will be mainly interested in the remaining two cases, i.e., when the Filippov 
solutions cross from $D^-$ into $D^+$ or vice versa. Indeed  in the cases where 
$f^+_N$ and $f^-_N$ share their sign, i.e., $f^\pm_N>0$  (Figure 
\ref{fig:unique1}) and $f^\pm_N<0$ (Figure \ref{fig:unique2}) solutions pass from 
$D^-$ into $D^+$ and from $D^+$ to $D^-$, respectively. In this case one speaks 
of \emph{transversally crossing trajectories}. More precisely the following 
criterion holds:
\bigskip

\begin{minipage}{.48\textwidth} \centering
  \begin{tikzpicture}
    \draw [very thick] (0,0) .. controls (1,1) and (3,0) .. 
    (5,3) node[align=right,below] {\,\,$N$};  
    \draw [very thick] (0,0) node[align=left,above] {$D^+$\qquad};
    \draw [very thick] (0,0) node[align=left,below] {$D^-$\qquad};
    \draw [->] (2.5,.85) -- (2.3,1.4) node[align=left,below] {$\vec n$\quad};
    
    \draw [->, red] (1.5,1) -- (1.7,2) node[align=right,above]{$f^+$};
    \draw [->, red] (3.5,2) -- (3.2,2.7);
    \draw [->, blue] (3,-.2) -- (4,-.15) node[align=right,below] 
    {\quad\,\,$f^-$};
    \draw [->, blue] (4,1) 
    -- (5,1.2);
    \draw [->, red] (.5,.35) -- (.7,1.2) ;
    \draw [->, blue] (.5,.35) -- (1.7,-.3);
    \draw [->, red] (4.3,2.1) -- (4.0,3);
    \draw [->,blue] (4.3,2.1) -- (5.4,2);
  \end{tikzpicture}
  
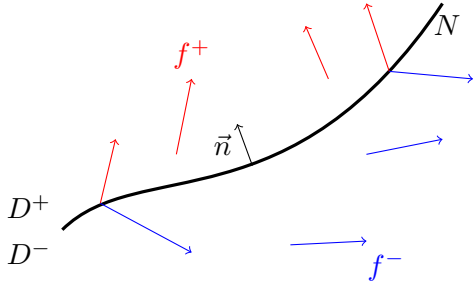
\captionof{figure}{repulsive trajectories} \label{fig:unique3}
\end{minipage}
\hfill
\begin{minipage}{.48\textwidth} \centering
  \begin{tikzpicture}
    \draw [very thick] (0,0) .. controls (1,1) and (3,0) .. 
    (5,3) node[align=right,below] {\,\,$N$};  
    \draw [very thick] (0,0) node[align=left,above] {$D^+$\qquad};
    \draw [very thick] (0,0) node[align=left,below] {$D^-$\qquad};
    \draw [->] (2.5,.85) -- (2.3,1.4) node[align=left,below] {$\vec n$\quad};
    
    \draw [->, red] (.5,1.7) -- (1.4,1.5) node[align=right,above]{$f^+$};
    \draw [->, red] (3,2) -- (3.7,1.9);
    \draw [->, blue] (2.5,-.5) node[align=right,above] {\quad\,\,$f^-$} -- 
    (2.3,.2) ;
    \draw [->, blue] (4,0) 
    -- (4.2,.7);
    \draw [->, red] (.5,.35) -- (2,0) ;
    \draw [->, blue] (.5,.35) -- (.3,1.2);
    \draw [->, red] (4.3,2.1) -- (5.3,1.5);
    \draw [->,blue] (4.3,2.1) -- (4.4,2.9);
  \end{tikzpicture}
  
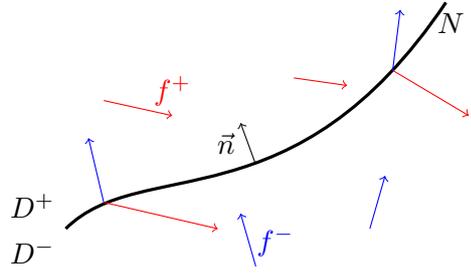
\captionof{figure}{sliding motion} 
  \label{fig:unique4}
\end{minipage}

\begin{minipage}{.48\textwidth} \centering
  \begin{tikzpicture}
    \draw [very thick] (0,0) .. controls (1,1) and (3,0) .. 
    (5,3) node[align=right,below] {\,\,$N$};  
    \draw [very thick] (0,0) node[align=left,above] {$D^+$\qquad};
    \draw [very thick] (0,0) node[align=left,below] {$D^-$\qquad};
    \draw [->] (2.5,.85) -- (2.3,1.4) node[align=left,below] {$\vec n$\quad};
    
    \draw [->, red] (1.5,1) -- (1.7,2) node[align=right,above]{$f^+$};
    \draw [->, red] (3.5,2) -- (3.2,2.7);
    \draw [->, blue] (2,-.5) node[align=right,above] {\quad\,\,$f^-$} -- 
    (1.8,.2) ;
    \draw [->, blue] (4,0) 
    -- (4.2,.7);
    \draw [->, red] (.5,.35) -- (.7,1.2) ;
    \draw [->, blue] (.5,.35) -- (.3,1.2);
    \draw [->, red] (4.3,2.1) -- (4.0,3);
    \draw [->,blue] (4.3,2.1) -- (4.4,2.9);
  \end{tikzpicture}
  
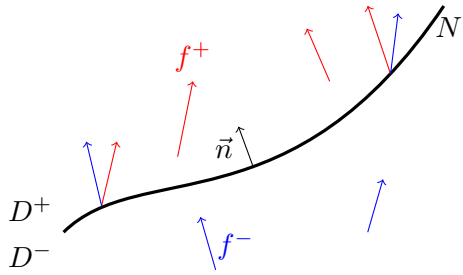
\captionof{figure}{(upward) trans\-ver\-sal\-ly crossing } \label{fig:unique1}
\end{minipage}
\hfill
\begin{minipage}{.48\textwidth} \centering
  \begin{tikzpicture}
    \draw [very thick] (0,0) .. controls (1,1) and (3,0) .. 
    (5,3) node[align=right,below] {\,\,$N$};  
    \draw [very thick] (0,0) node[align=left,above] {$D^+$\qquad};
    \draw [very thick] (0,0) node[align=left,below] {$D^-$\qquad};
    \draw [->] (2.5,.85) -- (2.3,1.4) node[align=left,below] {$\vec n$\quad};
    
    \draw [->, red] (.5,1.7) -- (1.4,1.5) node[align=right,above]{$f^+$};
    \draw [->, red] (3,2) -- (3.7,1.9);
    \draw [->, blue] (3,-.2) -- (4,-.15) node[align=right,below] 
    {\quad\,\,$f^-$};
    \draw [->, blue] (4,1) 
    -- (5,1.2);
    \draw [->, red] (.5,.35) -- (2,0) ;
    \draw [->, blue] (.5,.35) -- (1.7,-.3);
    \draw [->, red] (4.3,2.1) -- (5.3,1.5);
    \draw [->,blue] (4.3,2.1) -- (5.4,2);
  \end{tikzpicture}
  
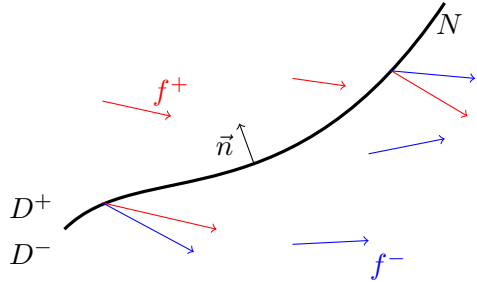
\captionof{figure}{(downward) trans\-ver\-sal\-ly crossing } 
  \label{fig:unique2}
\end{minipage}
\bigskip

\begin{cor}(Sufficient conditions for uniqueness \cite[Cor.\ 
10.1]{Fil:88}\label{cor:uni})
  Let $f$ be as above. On the region of the surface $N$ where ${f^+_N>0}$ and 
  ${f^-_N>0}$, Filippov solutions that reach $N$ from $D^-$ pass to 
  $D^+$ and hence uniqueness is not 
  violated. The analogous assertion holds for ${f^+_N, f^-_N<0}$ and 
  solutions passing from  $D^+$ to $D^-$.
\end{cor}

We will now apply Corollary \ref{cor:uni} to a geometric scenario that 
generically arises e.g.\ in `matched spacetimes'. Essentially the present 
discussion isolates  the abstract core of the one given in \cite[Sec.\ 
3.3]{PSSS:15} by neglecting the specific form of the spacetime metric 
used there. We start with a semi-Riemannian manifold $(M,g)$ with $g\in 
C^{0,1}$ and assume that $g$ is \emph{smooth off a $C^\infty$-hypersurface 
$\cN$} in the following sense: 
$M$ is the disjoint union of some open sets $\cD^+$, $\cD^-$ and their common 
boundary $\cN=\partial \cD^+=\partial \cD^-$ which we assume to be a smooth 
hypersurface of $M$.  The metric $g$ is smooth in $\cD^\pm$ up 
to the boundary, that is $g\in C^2(\cD^\pm\cup\cN)$. (Here $\cD^\pm\cup\cN$ are 
smooth manifolds with boundary.)

Now a (Filippov solution of the) geodesic (equation) 
in $\cD^\pm$ is a classical geodesic. Let us consider 
such a curve starting, say in $D^-$ (the case $D^+$ being analogous) and 
reaching $\cN$. Locally in coordinates $({\mathcal U},(x^1,\dots,x^n))$ we 
write $\cN$  as $\{x^1=0\}$ and we rewrite the geodesic equation for 
$\gamma(t)=(x^1(t),\dots,x^n(t))$ as a first  order system of the form
\begin{equation}\label{eq:gfo}
    \dot x^j=\hat x^j,\quad 
    \dot{\hat x}^j=-\Gamma^j_{\,km}(x)\,\hat x^k \hat x^m,
    \qquad (1\leq j\leq n)\,.
\end{equation}
Now the right hand side of \eqref{eq:gfo} is given by the 
$L^\infty_{\mbox{\scriptsize loc}}$-vector field 
$$f(x^1,\hat 
x^1,\dots,x^n,\hat x^n)=(\hat x^1,-\Gamma^1_{\,km}(x)\hat x^k\hat x^m,\dots, 
\hat x^n,-\Gamma^n_{\,km}(x)\hat x^k\hat x^m)
$$ 
defined on some open subset $U$ of $\R^{2n}$. On $U$ we  set $N:=\{x^1=0\}$ and 
$D^\pm:=\{\pm x^j>0\}$. Then the unit normal of $N$ pointing from $D^-$ to $D^+$ 
is $e_1$, i.e., the first standard unit vector. It follows that the projection 
of the limits of $f|_{D^\pm}$ on $N$ onto the normal coincide and are just given 
by $f^\pm_N=\hat x^1=\dot x^1$. So Corollary~\ref{cor:uni} applies if $\dot 
x^1(t_0)\not=0$ ($\dot x^1(t_0)>0$ in this case) where $t_0$ is the parameter 
value when $\gamma$ (first) hits $\cN$, i.e., $x^1(t_0)=0$. But this just means 
that $\gamma$ does hit $\cN$ transversally and we have the following result:

\begin{prop}(Sufficient conditions for uniqueness)\label{thm:scu}
  Let $(M,g)$ be a $C^{0,1}$-semi-Riemannian manifold with $g$ smooth off a 
  $C^\infty$-hypersurface\footnotemark[1]  $\cN$. Then the (Filippov) geodesics 
  starting in $M\setminus\cN$ that hit $\cN$ transversally are globally unique 
  and they cross from $\cD^-$ into $\cD^+$ or vice versa.  
  \footnotetext[1]{In the sense specified above.}
\end{prop}

This proposition especially applies if the hypersurface $\cN$ is \emph{totally 
geodesic} in the following sense: Every (Filippov) geodesic starting in $\cN$ 
tangentially to $\cN$ remains initially in $\cN$. Note that the notion of a 
totally geodesic submanifold in low regularity becomes somewhat subtle. In 
particular, in the situation at hand the second fundamental form will 
generically be not defined on all of $\cN$. 


Observe that the (Filippov) geodesics $\gamma$ crossing $\cN$ at $\gamma(t_0)$ 
coincide for all $t\not=t_0$ with the smooth geodesics of $\cD^\pm$. Hence by 
their $C^1$-property $g(\dot\gamma,\dot\gamma)$ is globally constant and these 
geodesics do have a causal character. Moreover they are extremal curves by the 
following argument which we only detail in the Riemannian case: Suppose there is 
a curve $\lambda$ connecting $\gamma(t_1)\in\cD^-$ with $\gamma(t_2)\in\cD^+$, 
which is shorter than $\gamma|_{[t_1,t_2]}$. Then set $t'=\sup\{t>t_1: 
\gamma(t)=\lambda(t)\}$ and suppose $t'<t_0$. By continuity 
$\gamma(t')=\lambda(t')$ and we choose a totally normal neighbourhood 
$U\subseteq\cD^-$ of $\gamma(t')$. Then $\lambda$ has to be minimizing in $U$ 
which contradicts the fact that $\lambda$ is not the radial geodesic $\gamma$ in 
$U$. So $\gamma(t)=\lambda(t)$ for all $t\not=t_0$ and hence everywhere by 
continuity.

Finally the above result suggests 
the idea to explicitly obtain the geodesics of $M$ by appropriately matching the 
geodesics of each `side' $\cD^\pm$ across $\cN$. 
We will discuss these matters in the closing section below.

\subsection{The $C^1$-matching of geodesics}
In this final section we discuss the matching of geodesics in locally Lipschitz 
semi-Riemannian manifolds with the metric smooth off a 
hypersurface $\cN$. Indeed such an approach has been frequently applied in the 
literature on impulsive gravitational waves, see e.g.\ 
\cite{FPV:88,PS:03,PS:10} and the references given in \cite[Sec.\ 
3.2]{PSSS:15}. The idea is to explicitly calculate the geodesics in 
$\cD^\pm$ which is often possible (only) 
in coordinates which do not extend to the `matching hypersurface' $\cN$. Then 
one matches the geodesics $\gamma^-$  of $\cD^-$ that hit $\cN$ `from below' to 
the geodesics $\gamma^+$ of $\cD^+$ that hit $\cN$ `from above' across $\cN$. 
Explicitly in coordinates which cover $\cN$ one sets 
$\gamma^-(t_0)=\gamma^+(t_0)$ and $\dot\gamma^-(t_0)=\dot\gamma^+(t_0)$ where 
$t_0$ is the parameter value where the respective geodesics hit $\cN$. Obviously 
one needs to involve the derivatives to obtain the correct number of equations 
to match all the data and this is why one refers to this approach as 
$C^1$-matching procedure.  

Often such calculations were done heuristically without supplying 
the necessary arguments which we collect here, cf.\ also \cite[Rem.\ 
4.1]{PSSS:15}. In fact the matching mathematically makes sense only
if the following facts on the geodesics of $M$ have been established:
\begin{enumerate}
  \item The geodesics reaching $\cN$ cross it (rather than 
  being reflected by or trapped into $\cN$).
  \item These geodesics are unique (rather than e.g.\ branching).
  \item These geodesics are at least of $C^1$-regularity.  
\end{enumerate}

Indeed we can extract all this necessary information from the discussion in the 
previous section to obtain the following result:
\begin{cor}(The $C^1$-matching)\label{cor:c1m}
  Let $(M,g)$ be a $C^{0,1}$-semi-Riemannian manifold with $g$ smooth off a 
  smooth hypersurface\footnotemark[2] $\cN$. 
  Let $\gamma^-:(a,b]\to\cD^-\cup\cN$ be a smooth geodesic with 
  $\gamma(b)\in\cN$ and $\dot\gamma(b)\not\in T_{\gamma(b)}\cN$.
  Then there exists $c>b$ and a unique (Filippov) geodesic $\gamma\colon 
  (a,c)\to M$ such that $\gamma|_{(a,b]}=\gamma^-$.
  \footnotetext[2]{In the sense specified above.}
\end{cor}

Now the $C^1$-matching of the data can be made explicitly by defining 
$\gamma^+:=\gamma|_{[b,c)}$. Indeed we then have $\gamma^-(b)=\gamma^+(b)$ and 
$\dot{\gamma}^-(b)=\dot{\gamma}^+(b)$. Also we remark that this procedure 
allows to derive the (Filippov) geodesics crossing $\cN$ simply by matching the 
smooth `background' geodesics on either side in a $C^1$-manner without the need 
to go into the details of Filippov's theory. 

This holds true even in case one needs to invoke the fact that $\cN$ is totally 
geodesic to rule out that geodesics hit $\cN$ tangentially. Indeed this fact 
can often be derived purely from knowledge of the `background 
geodesics' $\gamma^\pm$ and the $C^1$-regularity of the (Filippov) geodesics cf.\ \cite[Sec.\ 
3.6]{PSSS:15}. 

Finally in case $\cN$ fails to be totally geodesic (as is the case for 
all classes of expanding impulsive gravitational waves) one might still 
make use of Proposition \ref{thm:scu} and Corollary \ref{cor:c1m} (and hence 
establish the $C^{1,1}$-matching) by showing `by hand' that  the `background 
geodesics' do meet $\cN$ transversally, cf.\ \cite[Sec.\ 3.3]{PSSS:16}.

\section{Outlook and open problems}

We conclude with listing some open question and discuss further lines of research. 

\begin{itemize}
 \item Riemannian du Bois-Reymond-trick for Lipschitz continuous metrics\\
 Let $g$ be a Lipschitz continuous Riemannian metric and let $\gamma$ be a Lipschitz continuous minimizer. Is there a way 
to see that $\gamma$ has to be $C^2$ and that it satisfies the geodesic 
equations in the sense of Filippov?

 \item Lorentzian du Bois-Reymond trick\\
 Let $g$ be a $C^1$ Lorentzian metric and let $\gamma$ be a Lipschitz continuous maximizer between timelike related points. 
Does $\gamma$ have to be timelike? This would imply that one could apply the du Bois-Reymond trick to get that $\gamma$ is 
$C^2$ and that it satisfies the geodesic equations.

 \item Regularity of maximal causal curves\\
 Is there an analogue of the result by Lytchak and Yaman \cite{LY:06} for Lorentzian H\"older or Lipschitz continuous 
metrics? Even for $C^1$ metrics the regularity of maximal curves is unclear (cf.\ the point above).

 \item Causal character of maximizing causal curves\\
 What is the minimal regularity of a Lorentzian metric to ensure that maximal causal curves have a causal character? By 
Example \ref{ex:ks} the metric has to be at least Lipschitz 
continuous\footnote{Note added in proof: In \cite{GL:17} it was meanwhile shown 
that $g\in C^{0,1}$ is actually sufficient.}.

 \item Properties of Filippov geodesics\\
 Can one say more about Filippov geodesics in Riemannian and Lorentzian signature? Do they have to be minimizing or 
maximizing, respectively, in some sense? Observe that it is not expected that 
Lorentzian Filippov geodesics have a causal 
character.
\end{itemize}

\ack
The authors are very grateful for the hospitality during the conference 
in Florence and want to especially thank Ettore Minguzzi for the friendly 
organization of that meeting. We also thank Michael Kunzinger,
Ji{\v r}{\'\i} Podolsk{\'y} and Robert {\v{S}}varc for constantly sharing 
their expertise. This work was supported by the Austrian Science 
Fund FWF, grants P28770 and P26859.

\section*{References}
\bibliographystyle{abbrv}
\bibliography{fproc}

\end{document}